\documentstyle[12pt,epsfig,amssymb]{article}

\begin{document}
\newcommand{\beq}{\begin{equation}}\newcommand{\eeq}{\end{equation}}
\newcommand{\barr}{\begin{eqnarray}}\newcommand{\earr}{\end{eqnarray}}

\newcommand{\andy}[1]{ }

\def\ask{\marginpar{?? ask:  \hfill}}
\def\fin{\marginpar{fill in ... \hfill}}
\def\note{\marginpar{note \hfill}}
\def\check{\marginpar{check \hfill}}
\def\discuss{\marginpar{discuss \hfill}}

\newcommand{\bm}[1]{\mbox{\boldmath $#1$}}
\newcommand{\bmsub}[1]{\mbox{\boldmath\scriptsize $#1$}}

\newcommand{\ket}[1]{| {#1} \rangle}
\newcommand{\bra}[1]{\langle {#1} |}
\def\wtilde{\widetilde}
\def\Ord{\mbox{O}}
\def\As{{\cal A}}
\def\cH{{\cal H}}
\def\cN{{\cal N}}
\def\cP{{\cal P}}
\def\cR{{\cal R}}
\def\cZ{{\cal Z}}
\renewcommand{\Re}{\mbox{Re}}
\renewcommand{\Im}{\mbox{\rm Im}}

\def\coltwovector#1#2{\left({#1\atop#2}\right)}
\def\upp{\coltwovector10}    \def\downn{\coltwovector01}


\begin{titlepage}
\begin{flushright}
\today \\
\end{flushright}
\vspace{.5cm}
\begin{center}
{\LARGE Quantum Zeno effects with ``pulsed" and ``continuous"
measurements

}

\quad

{\large P. Facchi\(^{(1)}\) and S. Pascazio\(^{(2)}\)
\\
           \quad \\
$^{(1)}$Atominstitut der \"Osterreichischen Universit\"aten,
Stadionallee
2, A-1020, Wien, Austria \\
$^{(2)}$Dipartimento di Fisica, Universit\`a di Bari \\
     and Istituto Nazionale di Fisica Nucleare, Sezione di Bari \\
 I-70126 Bari, Italy \\

}

\vspace*{.5cm}

{\small\bf Abstract}\\ \end{center}

{\small The dynamics of a quantum system undergoing measurements
is investigated. Depending on the features of the interaction
Hamiltonian, the decay can be slowed (quantum Zeno effect) or
accelerated (inverse quantum Zeno effect), by changing the time
interval between successive (pulsed) measurements or,
alternatively, by varying the ``strength" of the (continuous)
measurement.}

\end{titlepage}

\newpage

\setcounter{equation}{0}
\section{Introduction }
 \label{sec-introd}
 \andy{intro}

Frequent measurement can slow the time evolution of a quantum
system, hindering transitions to states different from the initial
one \cite{Beskow,Misra}. This phenomenon is commonly known as the
quantum Zeno effect (QZE) and has been experimentally tested on
oscillating systems, characterized by a finite Poincar\'e time
\cite{Cook88,Nagels,Cookdisc}. However, new phenomena occur when one
considers unstable systems, whose Poincar\'e time is infinite
\cite{Bernardini93}: in particular, other regimes become possible,
in which measurement {\em accelerates} the dynamical evolution,
giving rise to an {\em inverse} quantum Zeno effect (IZE)
\cite{Kofman,Pascazio96,downconv,PFFP}.

 The study of Zeno effects for {\em
bona fide} unstable systems is a more complicated problem, because
it requires the use of quantum field theoretical techniques and in
particular a thorough understanding of the Weisskopf-Wigner
approximation \cite{Gamow28} and the Fermi ``golden" rule
\cite{Fermi}: for an unstable system the form factors of the
interaction play a fundamental role and determine the occurrence
of a Zeno or an inverse Zeno regime, depending of the physical
parameters describing the system. The occurrence of new regimes is
relevant from an experimental perspective, in view of the recent
observation of non-exponential decay (leakage through a confining
potential) at short times
\cite{Wilkinson}.
The general features of the quantum mechanical evolution law at
short and long times are summarized in \cite{review,review1}.

The usual approach to QZE and IZE makes use of ``pulsed"
observations of the quantum state. However, one obtains
essentially the same effects by performing a ``continuous"
observation of the quantum state, e.g.\ by means of an intense
field, that plays the role of external, ``measuring" apparatus.
Although this idea has been revived only recently
\cite{Mihokova97,Schulman98}, it is contained, in embryo, in
earlier papers \cite{Kraus81}.

We will analyze the transition from Zeno to inverse Zeno in the
context outlined above, by considering several examples, and will
then look at the temporal behavior of a three-level system (such
as an atom or a molecule) shined by an intense laser field
\cite{PFFP}: for physically sensible values of the intensity of
the laser, the decay can be {\em enhanced}, giving rise to an
inverse quantum Zeno effect.

\setcounter{equation}{0}
\section{Pulsed measurements}
\label{sec-QZE2lev}
\andy{QZE2lev}

We first consider ``pulsed" measurements, as in the seminal
approaches (\cite{vonNeumann32,Beskow,Misra}). The complementary
notion of ``continuous measurement" will be discussed in Sec.\
\ref{sec-QZEcont}.

\begin{figure}[t]
\begin{center}
\epsfig{file=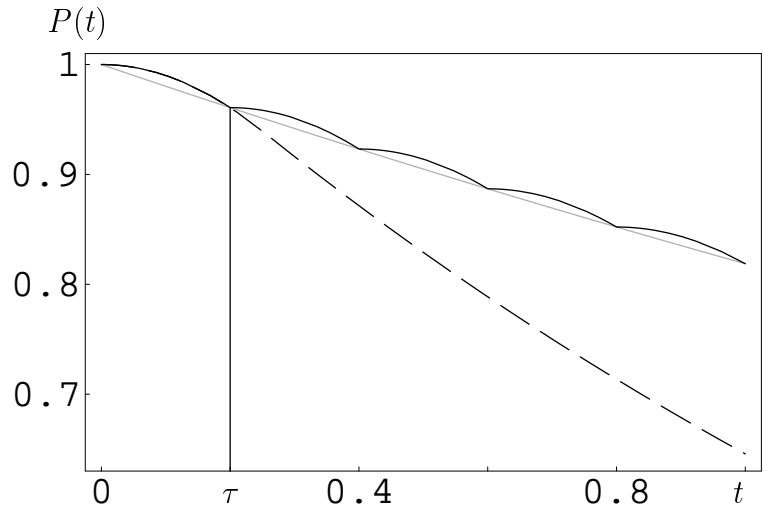,width=10cm}
\end{center}
\caption{Quantum Zeno effect due to ``pulsed" measurements.
The dashed (full) line is the survival probability without (with)
measurements. ($t$ in arbitrary units.)}
\label{fig:zenoevol}
\end{figure}

Let $H$ be the total Hamiltonian of a quantum system. The
survival amplitude and probability of the system in state
$|\psi_0\rangle$ are
\andy{survampl,survpr}
\barr
\As(t) &=& \langle \psi_0|\psi_t\rangle = \langle
\psi_0|e^{-iHt}|\psi_0\rangle,
\label{eq:survampl} \\
P(t) &=& |\As (t)|^2 =|\langle \psi_0|e^{-iHt}|\psi_0\rangle |^2,
\label{eq:survpr}
\earr
respectively. An elementary expansion yields a quadratic behavior
at short times
\andy{naiedef}
\beq
P(t) \sim 1 - t^2/\tau_{\rm Z}^2 \sim e^{-t^2/\tau_{\rm Z}^2},
\qquad
\tau_{\rm Z}^{-2}
\equiv \langle\psi_0|H^2|\psi_0\rangle -
\langle\psi_0|H|\psi_0\rangle^2 .
\label{eq:naiedef}
\eeq
It is often claimed that $\tau_{\rm Z}$ (``Zeno time") yields a
quantitative estimate of the short-time behavior. This is {\em
misleading} in many situations: $\tau_{\rm Z}$ is only the
convexity of $P(t)$ in the origin. This is a central observation
and will be corroborated by several examples. See
\cite{PIO} for an exhaustive discussion.

\begin{figure}[t]
\begin{center}
\epsfig{file=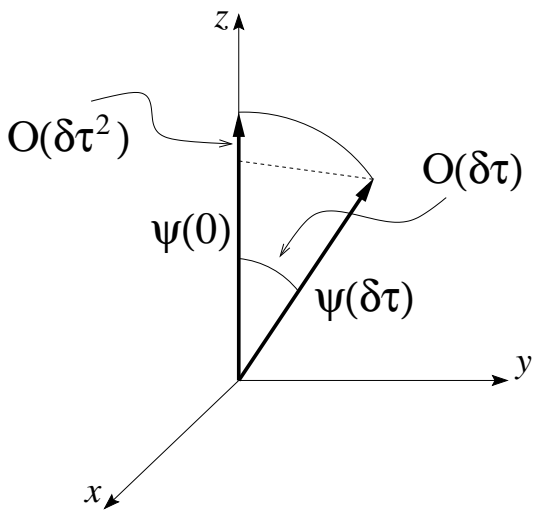,width=7cm}
\end{center}
\caption{Short-time evolution of phase and probability.}
\label{fig:phprob}
\end{figure}

Observe that if one divides the Hamiltonian into a free and an
interaction part
\beq
H=H_0 + H_{\rm I}
\eeq
and sets (for additional details and mathematical rigor, see
\cite{review,PIO,Nakazato97})
\beq
H_0\ket{\psi_0}=E_0\ket{\psi_0},\quad \bra{\psi_0}H_{\rm
I}\ket{\psi_0}=0,
\eeq
the Zeno time reads
\andy{ZenoHi}
\beq
\tau_{\rm Z}^{-2} = \langle\psi_0|H_{\rm I}^2|\psi_0\rangle
\label{eq:ZenoHi}
\eeq
and depends only on the off-diagonal part of the Hamiltonian.

In order to get QZE, we perform $N$ measurements at time intervals
$\tau$, to check whether the system is still in its initial state.
 The survival probability after the
measurements reads
\andy{survN}
\barr
P^{(N)}(t)&=&P(\tau)^N = P(t/N)^N\nonumber\\
& &\stackrel{N \; {\rm large}}{\longrightarrow}\left[ 1 -
(t/N\tau_{\rm Z})^2 \right]^N\sim
\exp(-t^2/N\tau_{\rm Z}^2)
\stackrel{N \rightarrow\infty}{\longrightarrow} 1 ,
 \label{eq:survN}
\earr
where $t=N\tau$ is the total duration of the experiment. The Zeno
evolution is pictorially represented in Figure \ref{fig:zenoevol}.

The QZE is a direct consequence of the Schr\"odinger equation that
yields quadratic behavior of the survival probability at short
times: in a short time $\delta\tau \sim 1/N $, the phase of the
wave function evolves like $\Ord (\delta\tau)$, while the
probability changes by $\Ord (\delta\tau^2)$, so that
\andy{survN3}
\beq
P^{(N)}(t) \sim \left[ 1 - \Ord (1/N^2)\right]^N \stackrel{N
\rightarrow \infty}{\longrightarrow}1.
 \label{eq:survN3}
\eeq
This is sketched in Figure \ref{fig:phprob} and is a very general
feature of the Schr\"odinger equation. As a matter of fact, many
other fundamental physical equations share the same property.

\setcounter{equation}{0}
\section{A theorem: from quantum Zeno to inverse quantum Zeno}
\label{sec-QZEinv}
\andy{QZEinv}

We rewrite (\ref{eq:survN}) as
\andy{survN0}
\beq
P^{(N)}(t)=P(\tau)^N=\exp(N\log P(\tau))= \exp(-\gamma_{\rm
eff}(\tau) t) ,
 \label{eq:survN0}
\eeq
where we introduced an effective decay rate \cite{FNP00,PIO}
\andy{effgamma0}
\beq
\gamma_{\rm eff}(\tau) \equiv -\frac{1}{\tau}\log P(\tau)
= -\frac{2}{\tau}\log |\As (\tau)| =-\frac{2}{\tau}\Re [\log
\As(\tau)].
 \label{eq:effgamma0}
\eeq
For instance, for times $\tau$ such that the quadratic behavior
(\ref{eq:naiedef}) is valid with good approximation, one easily
checks that
\andy{effgamma}
\beq
\gamma_{\rm eff}(\tau) \sim \tau/\tau_{\rm Z}^2, \qquad
(\tau\to 0)
 \label{eq:effgamma}
\eeq
is a linear function of $\tau$. We now concentrate our attention
on a truly unstable system, with ``natural" decay rate $\gamma$,
given by the Fermi ``golden" rule
\cite{Fermi}. We ask whether it is possible to find a finite time $\tau^*$
such that
\andy{tstardef}
\beq
\gamma_{\rm eff}(\tau^*)=\gamma.
\label{eq:tstardef}
\eeq
If such a time exists, then by performing measurements at time
intervals $\tau^*$ the system decays according to its ``natural"
lifetime, as if no measurements were performed. The related
concept of ``jump" time was considered in \cite{Schulman97}. See
also \cite{review,SRM}.

\begin{figure}[t]
\begin{center}
\epsfig{file=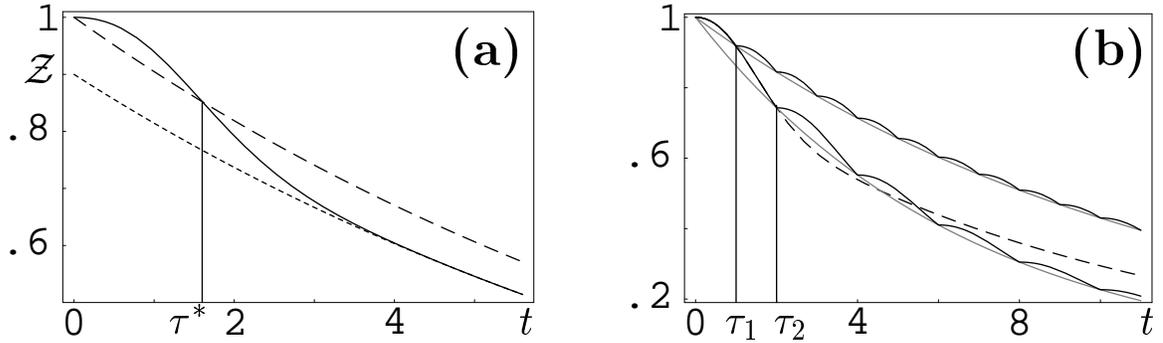,width=\textwidth}
\end{center}
\caption{$\cZ<1$. (a) Determination of $\tau^*$.
(b) Quantum Zeno vs inverse quantum Zeno (``Heraclitus") effect.}
\label{fig:gtau}
\end{figure}

By Eqs.\ (\ref{eq:tstardef}) and (\ref{eq:effgamma0}) one gets
\beq
P(\tau^*)=e^{-\gamma \tau^*},
\eeq
i.e., $\tau^*$ is the intersection between the curves $P(t)$ and
$e^{-\gamma t}$. In the situation depicted in Figure
\ref{fig:gtau}(a) such a time $\tau^*$ exists: the full line is
the survival probability and the dashed line the exponential
$e^{-\gamma t}$ [the dotted line is the asymptotic exponential
$\cZ e^{-\gamma t}$, see (\ref{eq:psurv}) in the following]. The
physical meaning of $\tau^*$ can be understood by looking at
Figure \ref{fig:gtau}(b), where the dashed line represents a
typical behavior of the survival probability $P(t)$ when no
measurement is performed: the short-time Zeno region is followed
by an approximately exponential decay with a natural decay rate
$\gamma$. When measurements are performed at time intervals
$\tau$, we get the effective decay rate $\gamma_{\rm eff}(\tau)$.
The full lines represent the survival probabilities and the
dotted lines their exponential interpolations, according to
(\ref{eq:survN0}). If $\tau=\tau_1<\tau^*$ one obtains QZE. {\em
Vice versa}, if $\tau=\tau_2>\tau^*$, one obtains IZE. If
$\tau=\tau^*$ one recovers the natural lifetime, according to
(\ref{eq:tstardef}): in this sense, $\tau^*$ can be viewed as a
{\em transition time} from a quantum Zeno to an inverse quantum
Zeno regime. Paraphrasing Misra and Sudarshan \cite{Misra} we can
say that $\tau^*$ determines the transition from Zeno (who argued
that a sped arrow, if observed, does not move) to Heraclitus (who
replied that everything flows).
\begin{figure}[t]
\begin{center}
\epsfig{file=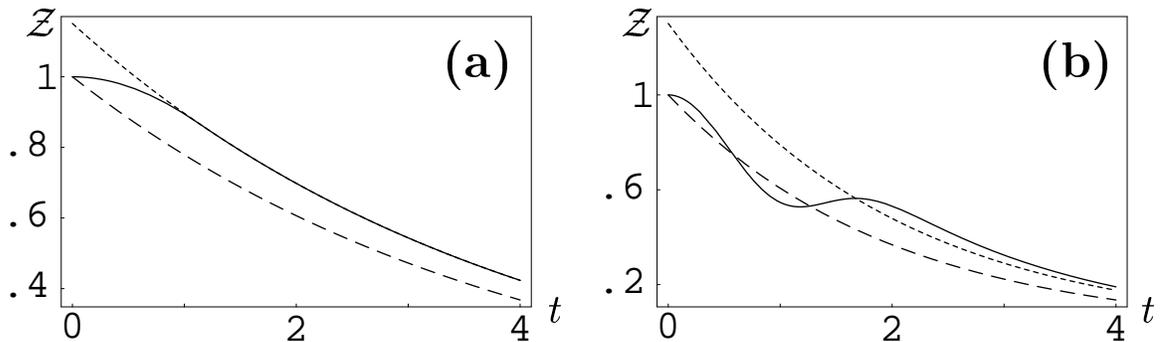,width=\textwidth}
\end{center}
\caption{$\cZ> 1$. The full line is the survival
probability, the dashed line the renormalized exponential
$e^{-\gamma t}$ and the dotted line the asymptotic exponential
$\cZ e^{-\gamma t}$. (a) If $P(t)$ and $e^{-\gamma t}$ do not
intersect, a finite solution $\tau^*$ does not exist. (b) If
$P(t)$ and $e^{-\gamma t}$ intersect, a finite solution $\tau^*$
exists. (In this case there are always at least two
intersections.) }
\label{fig:Zren}
\end{figure}

We stress that it is not always possible to determine $\tau^*$:
Eq.\ (\ref{eq:tstardef}) may have no finite solutions. Indeed,
for sufficiently long times the survival probability of an
unstable system reads with very good approximation
\andy{Psurv}
\beq
\label{eq:psurv}
P(t)= |\As(t)|^2 \simeq  \cZ e^{-\gamma t} ,
\eeq
where $\cZ$, the intersection of the asymptotic exponential with
the $t=0$ axis, is the wave function renormalization and is given
by the square modulus of the residue of the pole of the
propagator \cite{review,PIO,FNP00}.

A sufficient condition for the existence of a solution $\tau^*$
of Eq.\ (\ref{eq:tstardef}) is that $\cZ<1$. This is easily
proved by graphical inspection. The case $\cZ<1$ is shown in
Figure \ref{fig:gtau}(a): $P(t)$ and $e^{-\gamma t}$ must
intersect, since according to (\ref{eq:psurv}), $P(t) \sim \cZ
e^{-\gamma t}$ for large $t$, and a finite solution $\tau^*$ can
always be found. The other case, $\cZ > 1$, is shown in Figure
\ref{fig:Zren}: a solution may or may not exist, depending on the
features of the model investigated. There are also situations
(e.g., oscillatory systems, whose Poincar\'e time is finite) where
$\gamma$ and $\cZ$ cannot be defined
\cite{PIO}. The (existence of a) transition from Zeno to inverse Zeno
is therefore a complex, model-dependent problem, that requires
careful investigation. The theorem shown in this section gives a
simple criterion for the occurrence of such a transition.

\setcounter{equation}{0}
\section{Continuous measurement}
\label{sec-QZEcont}
\andy{QZEcont}

We now introduce some alternative descriptions of a measurement
process and discuss the notion of continuous measurement. This is
to be contrasted with the idea of pulsed measurements, discussed
in Section \ref{sec-QZE2lev} and hinging upon von Neumann's
projections \cite{vonNeumann32}.

Let us mimic the action of an external apparatus by the non
Hermitian Hamiltonian
\andy{iV}
\beq
H_{\rm I} = \pmatrix{0 & \Omega \cr \Omega & -i2V} = -iV{\bf 1} +
\bm h \cdot \bm\sigma ,
\label{iV}  \qquad \bm h =(\Omega,0,iV)^T ,
\eeq
that yields Rabi oscillations of frequency $\Omega$, but at the
same time absorbs away the $|-\rangle=\downn$ component of the
quantum state, performing in this way a ``measurement." Due to the
non Hermitian features of this description, probabilities are not
conserved: we are concentrating our attention only on the
$|+\rangle=\upp$ component.

\begin{figure}[t]
\begin{center}
\epsfig{file=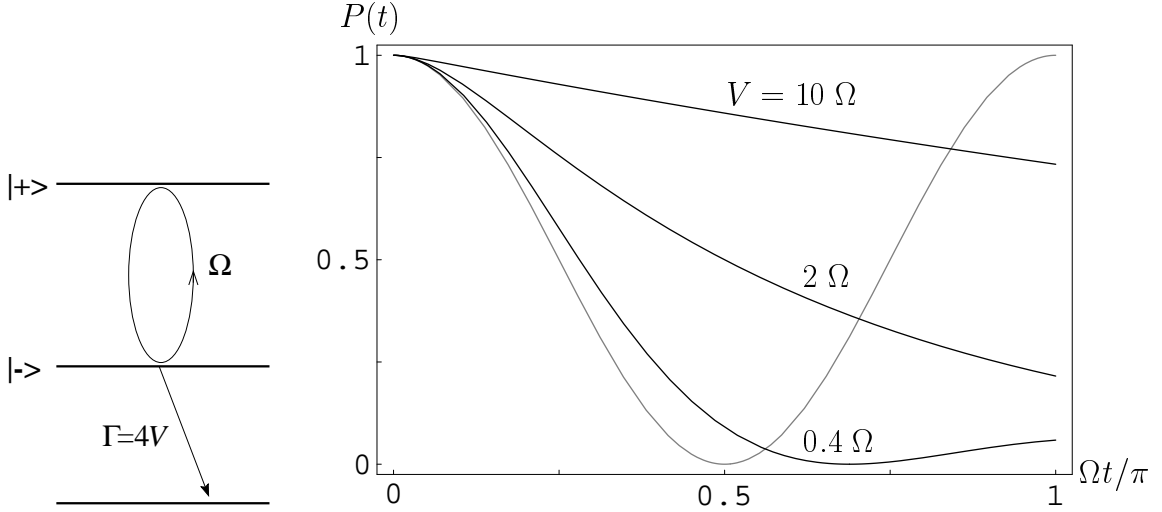,width=\textwidth}
\end{center}
\caption{Survival probability for a system undergoing Rabi
oscillations in presence of absorption ($V=0.4, 2, 10 \Omega$).
The gray line is the undisturbed evolution ($V=0$). }
\label{fig:zenoiV}
\end{figure}

An elementary calculation yields the evolution operator
\andy{eiVt}
\beq
e^{-iH_{\rm I}t} = e^{-Vt} \left[ \cosh (ht) -i \frac{\bm h\cdot
\bm \sigma}{h}\sinh (ht) \right],
\label{eiVt}
\eeq
where $h=\sqrt{V^2-\Omega^2}$ and we supposed $V>\Omega$. Let the
system be initially prepared in state $|\psi_0\rangle=|+\rangle$:
the survival amplitude reads
\andy{survamplV}
\barr
\As (t) &=& \langle \psi_0|e^{-iH_{\rm
I}t}|\psi_0\rangle\nonumber\\
&=& e^{-Vt} \left[ \cosh (\sqrt{V^2-\Omega^2}t) +
\frac{V}{\sqrt{V^2-\Omega^2}}\sinh (\sqrt{V^2-\Omega^2}t) \right].
\label{eq:survamplV}
\earr
The above results are exact and display some general aspects of
the quantum Zeno dynamics. The survival probability
$P(t)=|\As(t)|^2$ is shown in Figure \ref{fig:zenoiV} for
$V=0.4,2,10 \Omega$.
As expected, probability is (exponentially) absorbed away as $t
\to \infty$. However, for large $V (\gg\Omega)$ the survival
probability becomes
\andy{largeV}
\beq
P(t)\sim\left(1+\frac{\Omega^2}{2V}\right)
\exp\left(-\frac{\Omega^2}{V} t \right),
\label{eq:largeV}
\eeq
and the effective decay rate $\gamma_{\rm eff}(V)=\Omega^2/V$
becomes smaller, eventually halting the ``decay" of the initial
state and yielding an interesting example of QZE: a larger $V$
entails a more ``effective" measurement of the initial state.
Notice that the expansion (\ref{eq:largeV}) becomes valid very
quickly, on a time scale of order $V^{-1}$.

The non Hermitian Hamiltonian (\ref{iV}) ``summarizes" the
evolution engendered by a Hermitian Hamiltonian acting on a
larger Hilbert space, if attention is restricted to the subspace
spanned by $\{\ket{+}, \ket{-}\}$: indeed, let
\andy{hamflat}
\beq
H= \Omega(\ket{+}\bra{-}+\ket{-}\bra{+}) +\int d\omega\;\omega
\ket{\omega} \bra{\omega} +\sqrt{\frac{\Gamma}{2\pi}}\int
d\omega\;( \ket{-} \bra{\omega}+ \ket{\omega} \bra{-}) ,
\label{eq:hamflat}
\eeq
describing a two level system coupled to the photon field in the
rotating-wave approximation. By writing the state of the system
at time $t$ as
\andy{stateflat}
\beq
\ket{\psi_t}=\As(t)\ket{+}+y(t)\ket{-}+\int d\omega\;
z(\omega,t)\ket{\omega},
\label{eq:stateflat}
\eeq
a straightforward manipulation yields
\andy{subdynflat}
\barr
i\dot \As(t)&=& \Omega y(t), \nonumber\\
i\dot y(t)&=& -i \frac{\Gamma}{2} y + \Omega \As(t).
\label{eq:subdynflat}
\earr
These are the {\em same} equations of motion obtained from the
non Hermitian Hamiltonian
\andy{nonhermham}
\beq
H=\Omega(\ket{+}\bra{-}+\ket{-}\bra{+})-i\frac{\Gamma}{2}
\ket{-}\bra{-}.
\label{eq:nonhermham}
\eeq
which is the same as (\ref{iV}) when one sets $\Gamma=4V$. QZE is
obtained by increasing $\Gamma$ in
(\ref{eq:hamflat})-(\ref{eq:nonhermham}): a larger coupling to
the environment (photons) leads to a more effective ``continuous"
observation on the system (quicker response of the environment),
and as a consequence to a slower decay.

As one can see, the only effect of the continuum in
(\ref{eq:hamflat}) is the appearance of the imaginary frequency
$-i\Gamma/2$. This is ascribable to the ``flatness" of the
continuum [there is no form factor or frequency cutoff in the
last term of Eq.\ (\ref{eq:hamflat})], which yields a purely
exponential (Markovian) decay of $y(t)$.

The process described in this section can be viewed as a
``continuous" measurement performed on the initial state
$\ket{+}$: state $\ket{-}$ is indeed continuously monitored with a
response time $1/\Gamma$ and as soon as it becomes populated, it
is detected within a time $1/\Gamma$. The ``effectiveness"
$\Gamma=4 V$ of the observation can be compared to the frequency
$\tau^{-1}= (t/N)^{-1}$ of measurements in the ``pulsed"
formulation. Indeed, for large values of $\Gamma$ one gets from
Eq.\ (\ref{eq:largeV})
\andy{effgammaG}
\beq
\gamma_{\rm eff}(\Gamma)\sim\frac{4\Omega^2}{\Gamma}=\frac{4}{\tau_{\rm Z}^2
\Gamma}, \qquad (\Gamma\to\infty),
\label{eq:effgammaG}
\eeq
which, compared with Eq.\ (\ref{eq:effgamma}), yields an
interesting relation between continuous and pulsed measurements
\andy{contvspuls}
\beq
\Gamma\simeq 4/\tau=4 N/t.
\label{eq:contvspuls}
\eeq
In a Zeno context, the strength of the coupling can therefore be
viewed as an inverse characteristic time. The effective lifetime
$\gamma_{\rm eff}$ is then a function of the time interval $\tau$
between successive (pulsed) measurements or, alternatively, of
the strength $\Gamma$ of the (continuous) measurement. This also
leads to a novel definition of QZE (and IZE)
\cite{PIO}. The relation between ``pulsed" and ``continuous"
measurements in the context of QZE was first emphasized by
Schulman \cite{Schulman98}.

\setcounter{equation}{0}
\section{Continuous ``Rabi" measurement }
\label{sec-QZEcont1}
\andy{QZEcont1}

Let us now analyze a different situation, by coupling state
$|-\rangle$ to another state $|M\rangle$, that will play the role
of measuring apparatus. This will clarify that absorption and/or
probability leakage to the environment (like those investigated
in the previous section) are {\em not} fundamental requisites to
obtain QZE.

The (Hermitian) Hamiltonian is
\andy{ham3l}
\beq
H_{\rm I} = \Omega ( |+\rangle \langle -| + |-\rangle \langle +|)
+ K ( |-\rangle \langle M| + |M\rangle \langle -|) = \pmatrix{0 &
\Omega & 0\cr \Omega & 0 & K \cr 0 & K & 0},
\label{ham3l}
\eeq
where $K \in \mathbb{R}$ is the strength of the coupling to the
new level $M$ and
\andy{+-M}
\beq
\langle+ | = (1,0,0), \quad \langle -| = (0,1,0), \quad \langle
M| = (0,0,1) .
\label{+-M}
\eeq
This is probably the simplest way to include an ``external"
apparatus in our description: as soon as the system is in
$|-\rangle$ it undergoes Rabi oscillations to $|M\rangle$.
Similar examples were considered by other authors
\cite{Kraus81}. We expect level $|M\rangle$  to perform
better as a measuring apparatus when the strength $K$ of the
coupling increases.

\begin{figure}[t]
\begin{center}
\epsfig{file=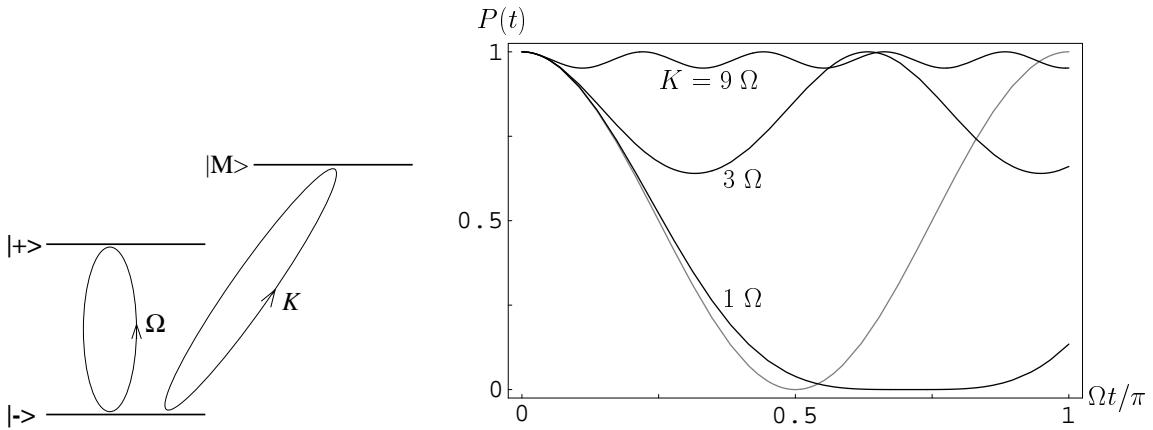,width=\textwidth}
\end{center}
\caption{Survival probability for a continuous Rabi ``measurement"
with $K=1,3,9\Omega$: quantum Zeno effect.}
\label{fig:zenocont}
\end{figure}

The initial state is $|\psi_0\rangle = |+\rangle$ and after a
straightforward calculation one obtains the survival probability
in state $|+\rangle$
\andy{sp3}
\beq
P(t)= \frac{1}{(K^2+\Omega^2)^2}\left[K^2+ \Omega^2
\cos(\sqrt{K^2+\Omega^2}t) \right]^2.
\label{sp3}
\eeq
This is shown in Figure \ref{fig:zenocont} for $K=1,3,9 \Omega$.
We notice that for large $K$ the state of the system does not
differ much from the initial state: as $K$ is increased, level
$|M\rangle$ performs a better ``observation" of the state of the
system, hindering transitions from $|+\rangle$ to $|-\rangle$.
This can be viewed as a QZE due to a ``continuous," yet Hermitian
observation performed by level $|M\rangle$.

This simple example enables one to make an important observation.
The Zeno time is easily computed and turns out to be much longer
than the Poincar\'e time $T_{\rm P}$ (we are assuming $K \gg
\Omega$)
\andy{tauz3}
\beq
\tau_{\rm Z} = \Omega^{-1} \gg T_{\rm P} = \Ord (K^{-1}).
\label{eq:tauz3}
\eeq
As a matter of fact, $\tau_{\rm Z}$ yields only the convexity of
the survival probability in the origin: in general, the short-time
quadratic evolution of the system is much shorter. This
contradicts many erroneous claims in the literature of the last
few years.

One can also obtain a relation similar to (\ref{eq:contvspuls}):
\andy{KvsN}
\beq
K \simeq \frac{\sqrt{2N}}{t}:
\label{eq:KvsN}
\eeq
again, strong coupling is equivalent to frequent measurements
\cite{PIO}.

A final comment is in order. All the situations analyzed in
Sections \ref{sec-QZEcont} and \ref{sec-QZEcont1} lead to QZE but
never to IZE. The reason for this is profound and lies in the
absence of the form factors of the interactions. The importance of
form factors and the role they play in this context is discussed
in \cite{PIO,FNP00} and will be clarified in the following
section.

\setcounter{equation}{0}
\section{Influence of form factors}
\label{sec-qmeas}
\andy{qmeas}

Consider the Hamiltonian
\andy{hamnotflat}
\beq
H_{\rm decay}= H_0+ H_{\rm I}=\omega_{\rm in}\ket{\rm in}\bra{{\rm
in}}+\sum_n \omega_n \ket{n} \bra{n} +\sum_n\left(
\phi_n\ket{n}\bra{{\rm in}} +
\phi^*_n \ket{{\rm in}} \bra{n}\right),
\label{eq:hamnotflat}
\eeq
with $\bra{n}n'\rangle=\delta_{nn'}$, $\bra{\rm in }\rm in
\rangle=1$ and $\bra{n}{\rm in}\rangle=0$, describing an initial state $\ket{\rm in}$
coupled to many discrete states (we will eventually consider the
continuum limit in order to get a decay process). The coupling is
not ``flat:" there is a form factor $\phi_n$. The state of the
system at time $t$ is
\andy{statek}
\beq
\ket{\psi_t}=\As(t)e^{-i\omega_{\rm in}t}\ket{\rm in}+\sum_n
z_n(t)\ket{n}
\label{eq:statek}
\eeq
and the Schr\"odinger equation, with the initial condition
$\ket{\psi_0}=\ket{\rm in}$, yields
 \andy{xdoteq1}
 \beq
\dot \As(t) = - \sum_n |\phi_n|^2\int_0^t ds\;
\exp [-i (\omega_n-\omega_{\rm in})s]
\As(t-s) .
 \label{eq:xdoteq1}
 \eeq
Observe that $\As(t)$ contains non-exponential contributions
\cite{Peres}.

\begin{figure}[t]
\centerline{\epsfig{file=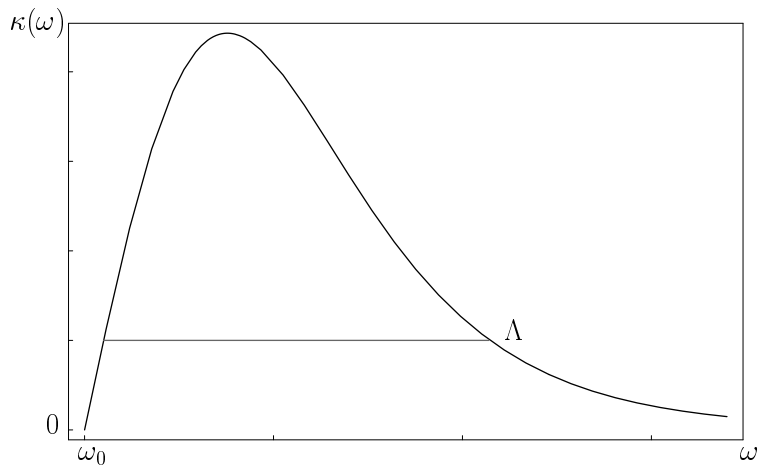, width=10cm}}
\caption{Typical behavior of the form factor $\kappa(\omega)$.}
\label{fig:matrixel}
\end{figure}

Let the (pulsed) measurements be performed at time intervals much
shorter than the natural lifetime, $\tau\ll 1/\gamma$. Since
$\tau$ is very small, $\As(\tau)\simeq
\As(0)=1$, and
\beq
\log \As(\tau)\sim \As(\tau)-1=\As(\tau)-\As(0)=\int_0^\tau dt\; \dot \As(t),
\eeq
so that the expression $\gamma_{\rm eff}(\tau)=-(2/\tau)\Re \{\log
\As(\tau)\}$ in (\ref{eq:effgamma0}) yields
\barr
\gamma_{\rm eff}(\tau)&\sim&\frac{2}{\tau}\;\Re
\sum_n |\phi_n|^2\int_0^\tau dt\int_0^t ds\;
e^{-i(\omega_n-\omega_{\rm in}) s} \As(t-s)\nonumber\\
&\sim& \frac{2}{\tau}\sum_n |\phi_n|^2\Re \int_0^\tau dt\int_0^t
ds\; \exp[-i(\omega_n-\omega_{\rm in}) s],
\earr
where we used again $\As(\tau)\simeq 1$. By performing the
integration
\beq
\gamma_{\rm eff}(\tau)
=\tau\sum_n |\phi_n|^2
\frac{\sin^2\left(\frac{(\omega_n-\omega_{\rm in})\tau}{2}\right)}
{\left(\frac{(\omega_n-\omega_{\rm in})\tau}{2}\right)^2}
\eeq
and introducing the spectral density function (form factor)
\andy{spd}
\beq
\kappa(\omega)=\sum_n \delta(\omega-\omega_n) |\phi_n|^2,
\label{eq:spd}
\eeq
whose support is the spectrum $\{\omega_n\}_n$, we get
\andy{sinc}
\beq
\gamma_{\rm eff}(\tau)=\tau\int_{\omega_0}^{\infty} d\omega\;
\kappa(\omega)
\frac{\sin^2\left(\frac{(\omega-\omega_{\rm in})\tau}{2}\right)}
{\left(\frac{(\omega-\omega_{\rm in})\tau}{2}\right)^2},
\label{eq:sinc}
\eeq
where $\omega_0$ is the ground state energy. This result was
first obtained by Kofman and Kurizki, last article in
\cite{Kofman}. We assume $\omega_0 <
\omega_{\rm in}$, in order to have a decaying initial state
$\ket{\rm in}$. Note that in the continuum limit the energies
$\omega_n$ form a continuum and the form factor $\kappa(\omega)$
becomes an ordinary function of $\omega$
\andy{spdcont}
\beq
\kappa(\omega)= \rho(\omega) |\varphi(\omega)|^2,
\label{eq:spdcont}
\eeq
where $\rho$ is the state density function and $\varphi$ the
rescaled matrix element. A typical behavior of the form factor is
shown in Figure \ref{fig:matrixel}: observe that
$\kappa(\omega<\omega_0)=0$ and
$\kappa(\omega>\omega_0+\Lambda)\simeq 0$, where $\Lambda$ is a
natural cutoff.

\begin{figure}[t]
\centerline{\epsfig{file=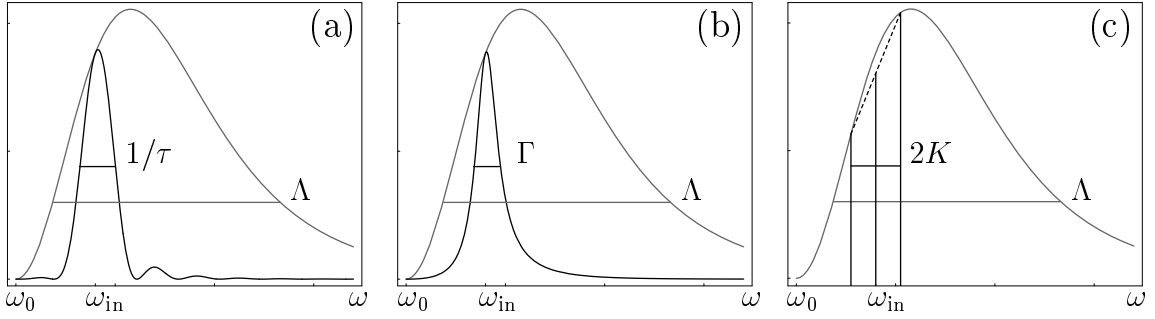,width=\textwidth}}
\caption{Form factor $\kappa(\omega)$ and detector ``response"
function for: (a) pulsed measurements, Eq.\ (\ref{eq:sinc}), with
detector response function ${\rm sinc}^2[(\omega-\omega_{\rm
in})\tau/2]$; (b) continuous measurement, Eq.\ (\ref{eq:gameffi}),
with detector response function $(\Gamma/2)^2/[(\omega-\omega_{\rm
in})^2+(\Gamma/2)^2]$; (c) ``Rabi" measurement, Eq.\
(\ref{eq:gameffrabi}), with detector response function
$[\delta(\omega-\omega_{\rm in}-K)+\delta(\omega-\omega_{\rm
in}+K)]$. The gray line is a typical form factor and the full line
represents the function in the integral, given by the appropriate
response function times the form factor.}
\label{fig:response}
\end{figure}

A pictorial representation of the integral (\ref{eq:sinc}) is
given in Figure \ref{fig:response}(a): the detector ``response"
function ${\rm sinc}^2[(\omega-\omega_{\rm in})\tau/2]$ [where
${\rm sinc} (x)\equiv\sin x/x$] is modulated by the form factor
$\kappa(\omega)$.

\begin{figure}[t]
\centerline{\epsfig{file=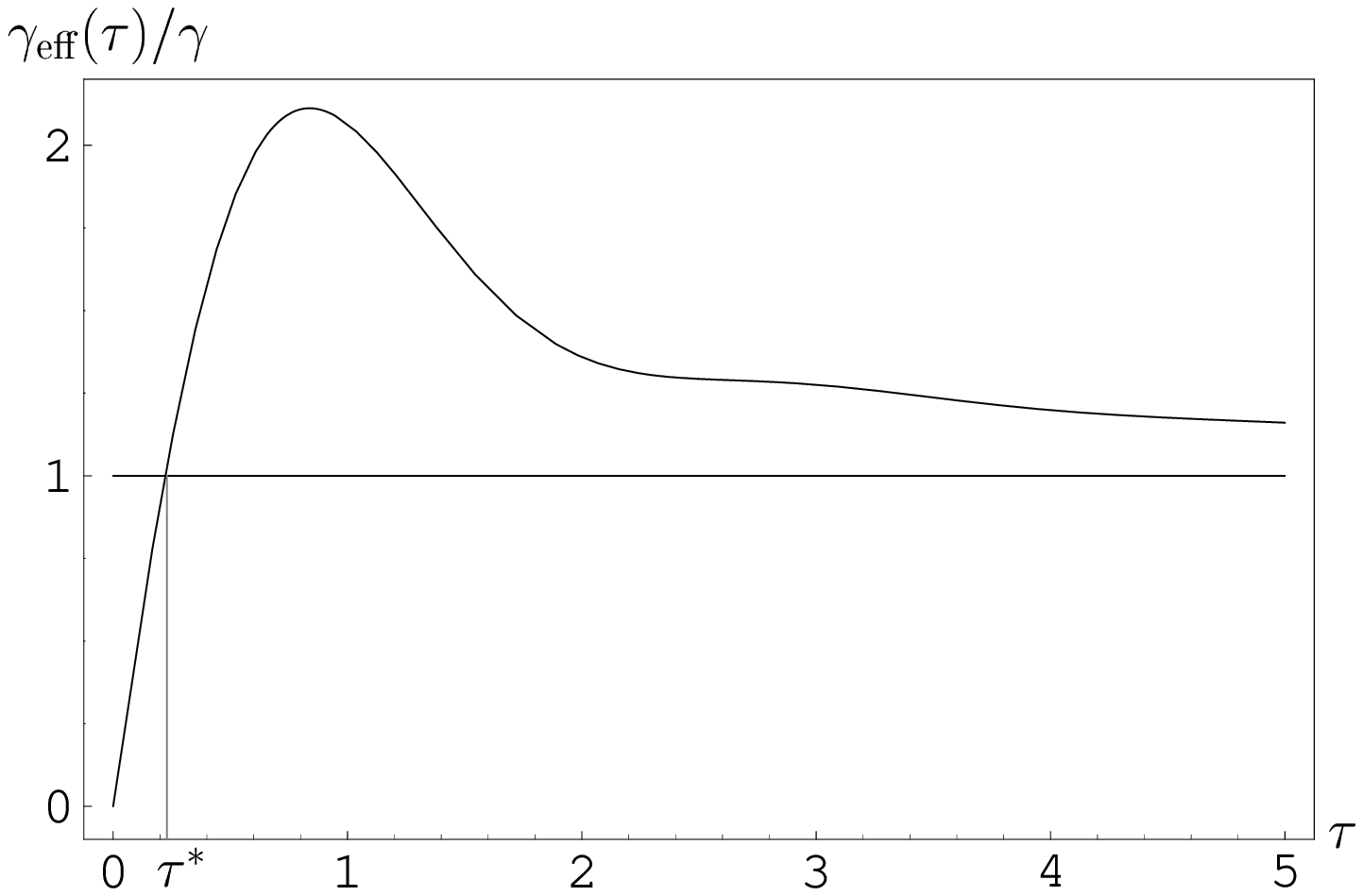,width=10cm}}
\caption{Effective decay rate $\gamma_{\rm eff}(\tau)$: typical
behavior for systems exhibiting a QZE-IZE transition. The
horizontal line shows the natural decay rate $\gamma$: its
intersection with $\gamma_{\rm eff}(\tau)$ yields the transition
time $\tau^*$, solution of Eq.\ (\ref{eq:tstardef}). Notice the
presence of a linear region for small values of $\tau$, according
to Eq.\ (\ref{eq:g0}). A Zeno (inverse Zeno) effect is obtained
for $\tau<\tau^*$ ($\tau>\tau^*$). For large $\tau$, the
asymptotic value of the curve is $\gamma$, as expected from Eq.\
(\ref{eq:ginf}).}
\label{fig:gammat}
\end{figure}

By using the asymptotic properties of sinc$^2(x)$, we get
\andy{ginf,g0}
\barr
\gamma_{\rm eff}(\tau) &\simeq& 2\pi\kappa(\omega_{\rm in})
=\gamma\qquad\mbox{for}\;\;\tau\gg 1/\Lambda,\label{eq:ginf}
\\
\gamma_{\rm eff}(\tau) &\simeq&
\tau\int_{\omega_0}^\infty d\omega\;\kappa(\omega)=
\frac{\tau}{\tau_{\rm Z}^2}
\qquad\mbox{for}\;\;\tau\ll 1/\Lambda,
\label{eq:g0}
\earr
where Eqs.\ (\ref{eq:spd}), (\ref{eq:hamnotflat}) and
(\ref{eq:ZenoHi}) were used to obtain
\beq
\int_{\omega_0}^\infty d\omega\;\kappa(\omega)=\sum_n
|\phi_n|^2=\bra{\rm in}H_{\rm I}^2\ket{\rm in}=\tau_{\rm Z}^{-2}.
\eeq
The first result (\ref{eq:ginf}) was to be expected: if the time
interval $\tau$ between successive pulsed measurements is long
enough, one recovers the ``natural" lifetime, computed according
to the Fermi ``golden" rule. The second result (\ref{eq:g0}) is
interesting and generalizes (\ref{eq:effgamma}), providing a
timescale $\Lambda^{-1}$ for its range of validity. Notice that
$\gamma_{\rm eff}$ is a linear function of $\tau$ in this range.
A typical behavior of $\gamma_{\rm eff}(\tau)$, for a form factor
like that in Figure \ref{fig:matrixel}, is shown in Figure
\ref{fig:gammat}: for
$\tau<\tau^*$, one has $\gamma_{\rm eff}<\gamma$ and therefore
QZE. {\em Viceversa}, for $\tau>\tau^*$, one has $\gamma_{\rm
eff}>\gamma$ and therefore IZE \cite{FNP00,PIO}.

The previous analysis and results are valid for pulsed
measurements. Let us consider now a continuous measurement
process. This is accomplished, for instance, by adding to
(\ref{eq:hamnotflat}) the following Hamiltonian
\andy{omomprime}
\beq
H_{\rm meas}(\Gamma) = \sqrt{\frac{\Gamma}{2\pi}} \sum_n\int
d\omega'\;(
\ket{n} \bra{n, \omega'}+
\ket{n,\omega'} \bra{n}) + \int
d\omega'\;\ket{\omega'}\bra{\omega'}:
\label{eq:omomprime}
\eeq
as soon as state $\ket{n}$ is populated, it is coupled to a boson
of frequency $\omega'$ (notice that the coupling has no form
factor). By following a reasoning identical to that of Section
\ref{sec-QZEcont}, one can show that the dynamics of the
Hamitonian (\ref{eq:hamnotflat}) $+$ (\ref{eq:omomprime}), in the
relevant subspace, is generated by
\andy{hamnotflatsub}
\beq
H=\omega_{\rm in}\ket{\rm in}\bra{{\rm in}}+\sum_n
\left(\omega_n-i\frac{\Gamma}{2}\right)
\ket{n} \bra{n} +\sum_n\left( \phi_n\ket{n}\bra{{\rm in}} +
\phi^*_n \ket{{\rm in}} \bra{n}\right)
\eeq
and a better continuous observation on the system is obtained by
increasing $\Gamma$, like in Section \ref{sec-QZEcont}. With the
substitution $\omega_n\to\omega_n-i\Gamma/2$, Eq.\
(\ref{eq:xdoteq1}) reads
\andy{AseqGamma}
\beq
\dot \As(t) = - \sum_n |\phi_n|^2\int_0^t ds\;
\exp [-i (\omega_n-\omega_{\rm in}-i\Gamma/2)s]
\As(t-s) .
 \label{eq:AseqGamma}
\eeq
For $t\gg1/\Lambda$ (and not extremely large), where $\Lambda$ is
the cutoff of the form factor (see Figure
\ref{fig:matrixel}), the decay is
exponential with very high accuracy and the decay rate is given by
\andy{gameffcont}
\beq
\gamma_{\rm eff}(\Gamma)=2\sum_n |\phi_n|^2\Re\int_0^\infty dt\;
e^{-i (\omega_n-\omega_{\rm in}-i\Gamma/2)t}=\frac{4}{\Gamma}
\sum_n |\phi_n|^2 \frac{\frac{\Gamma^2}{4}}{(\omega_n-\omega_{\rm
in})^2+\frac{\Gamma^2}{4}}.
\label{eq:gameffcont}
\eeq
In the continuum limit, (\ref{eq:gameffcont}) is expressed in
terms of the spectral density (\ref{eq:spdcont}):
\andy{gameffi}
\beq
\gamma_{\rm eff}(\Gamma)=\frac{4}{\Gamma}
\int_{\omega_0}^{\infty} d\omega\;
\kappa(\omega) \frac{\frac{\Gamma^2}{4}}{(\omega-\omega_{\rm
in})^2+\frac{\Gamma^2}{4}},
 \label{eq:gameffi}
\eeq
which is the analog of formula (\ref{eq:sinc}) for continuous
measurement. The function in the above integral is represented in
Figure \ref{fig:response}(b). The asymptotic values are
\andy{ginfcont,g0cont}
\barr
\gamma_{\rm eff}(\Gamma) &\simeq&\gamma\quad\mbox{for}
\quad\Gamma\ll \Lambda,
\label{eq:ginfcont}
\\
\gamma_{\rm eff}(\Gamma) &\simeq& \frac{4}{\tau_{\rm Z}^2\Gamma}
\quad\mbox{for}\quad\Gamma\gg \Lambda
\label{eq:g0cont}
\earr
and must be compared to (\ref{eq:ginf})-(\ref{eq:g0}). Once again,
the result (\ref{eq:ginfcont}) was to be expected: if the coupling
$\Gamma$ to the external field is weak enough (as compared to
$\Lambda$) the system decays according to its ``natural" lifetime,
computed according to the Fermi ``golden" rule. On the other hand,
Eq.\ (\ref{eq:g0cont}) is interesting and yields again
(\ref{eq:contvspuls}) [via (\ref{eq:g0})] in a more general
context. Furthermore, when $\Gamma$ varies between $0$ and
$\infty$, $\gamma_{\rm eff}(\Gamma)$ describes a curve very
similar to that relative to pulsed measurements and represented in
Figure \ref{fig:gammat}.

Finally, one can also consider the continuous ``Rabi" measurement
of Sec.\ \ref{sec-QZEcont1}, by adding to (\ref{eq:hamnotflat})
the following Hamiltonian
\andy{Rabicont}
\beq
H_{\rm meas}(K) = K \sum_n (
\ket{n} \bra{n, M}+
\ket{n,M} \bra{n}) + \sum_n \omega_n
\ket{n,M}\bra{n,M},
\label{eq:Rabicont}
\eeq
with $\bra{n,M}n',M\rangle=\delta_{n n'}$, $\bra{n}n',M\rangle=0$
and $\bra{\rm in}n,M\rangle=0$: as soon as state $\ket{n}$ is
populated, it undergoes Rabi oscillations to the orthogonal state
$\ket{n,M}$. The Schr\"odinger equation yields
 \andy{Rabixdoteq1}
 \beq
\dot \As(t) = - \sum_n |\phi_n|^2\int_0^t ds\; \exp [-i (\omega_n-\omega_{\rm
in})s]\cos (K s)\;
\As(t-s) ,
 \label{eq:Rabixdoteq1}
 \eeq
which is to be compared with Eqs.\ (\ref{eq:xdoteq1}) and
(\ref{eq:AseqGamma}). The exponential decay rate reads now
\barr
\gamma_{\rm eff}(K)&=&2\sum_n |\phi_n|^2\Re\int_0^\infty dt\;
e^{-i (\omega_n-\omega_{\rm in})t}\cos(K t)\nonumber\\
&=&\pi \sum_n |\phi_n|^2 \left[ \delta(\omega_n-\omega_{\rm
in}-K)+\delta(\omega_n-\omega_{\rm in}+K)\right]
\earr
and is expressed in terms of the spectral density function
\andy{gameffrabi}
\barr
\gamma_{\rm eff}(K)&=&
 \pi\int_{\omega_0}^{\infty} d\omega\;
\kappa(\omega) \left[ \delta(\omega-\omega_{\rm
in}-K)+\delta(\omega-\omega_{\rm in}+K)\right]\nonumber\\
& =&\pi\left[\kappa(\omega_{\rm in}+K)+\kappa(\omega_{\rm
in}-K)\right].
\label{eq:gameffrabi}
\earr
The decay rate is the arithmetic mean between two ``free" decay
rates, from initial states with shifted initial energies
$\omega_{\rm in}\pm K$, into the same continuum \cite{PFFP}. This
is shown in Figure \ref{fig:response}(c). The asymptotic values
read
\barr
\gamma_{\rm eff}(K) &\simeq&\gamma\quad\mbox{for}\quad K\ll \Lambda,
\\
\gamma_{\rm eff}(K) &=& \pi\kappa(\omega_{\rm in}+K)\quad\mbox{for}\quad K\gg\Lambda,
\earr
The physical meaning of the first expression is apparent. Note
that the second equation is already exact for $K>\omega_{\rm
in}-\omega_0$, because $\kappa(\omega)$ vanishes identically for
$\omega<\omega_0$. Note also that for $K\to\infty$ (infinitely
strong Rabi measurement) $\gamma_{\rm eff}(K)$ tends to zero
exactly like the form factor (QZE).

In the next section we will consider a physical situation in
which a decaying system is actually ``observed" via a Rabi
oscillation and will see that for sensible values of the
parameters the decay is enhanced (IZE).

\setcounter{equation}{0}
\section{Three-level system in a laser field}
 \label{sec-QED}
 \andy{QED}

\begin{figure}[t]
\begin{center}
\epsfig{file=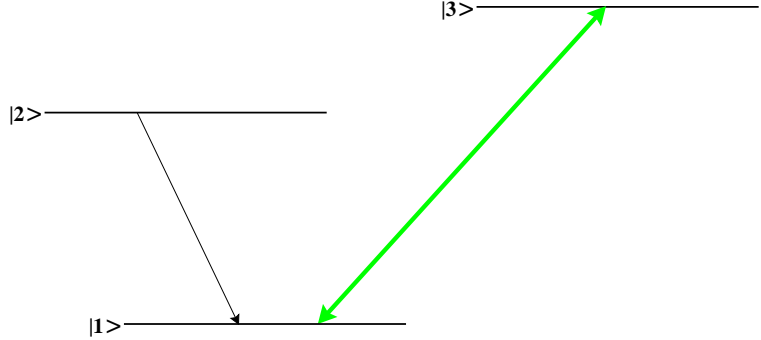,width=10cm}
\end{center}
\caption{Level configuration}
\label{fig:fig0}
\end{figure}

We now analyze a realistic situation in which a continuous
observation performed by a laser field leads to an inverse
quantum Zeno effect. We look at the temporal behavior of a
three-level system (such as an atom or a molecule), where level
$|1\rangle$ is the ground state and levels $|2\rangle$,
$|3\rangle$ are two excited states \cite{Mihokova97}. See Figure
\ref{fig:fig0}. The system is initially prepared in level
$|2\rangle$  and if it follows its natural evolution, it will
decay to level $|1\rangle$. The decay will be (approximately)
exponential and characterized by a certain lifetime, that can be
calculated from the Fermi ``golden" rule. But if one shines on the
system an intense laser field, tuned at the transition frequency
3-1, the decay is {\em enhanced} (inverse quantum Zeno effect)
\cite{PFFP}.

We consider the Hamiltonian ($\hbar=c=1$)
\andy{ondarothamdip6}
\barr
H &=& H_{0}+H_{\rm int}\nonumber\\
&=& \omega_0|2\rangle\langle 2|+\Omega_0|3\rangle\langle 3|
+\sum_{\bmsub k,\lambda}\omega_k a^\dagger_{\bmsub{k}\lambda}
a_{\bmsub{k}\lambda} +\sum_{\bmsub k,\lambda}\left(\phi_{\bmsub
k\lambda} a_{\bmsub k\lambda}^\dagger|1\rangle\langle2|
+\phi_{\bmsub k\lambda}^* a_{\bmsub
k\lambda}|2\rangle\langle1|\right)\nonumber\\ & &+\sum_{\bmsub
k,\lambda}\left(\Phi_{\bmsub k\lambda} a_{\bmsub
k\lambda}^\dagger|1\rangle\langle3| +\Phi_{\bmsub k\lambda}^*
a_{\bmsub k\lambda}|3\rangle\langle1|\right),
\label{eq:ondarothamdip6}
\earr
where the first two terms are the free Hamiltonian of the 3-level
atom (whose states $|i\rangle$ $(i=1,2,3)$ have energies $E_1=0$,
$\omega_0=E_2-E_1>0$, $\Omega_0=E_3-E_1>0$), the third term is the
free Hamiltonian of the EM field ($\lambda$ being a discrete
index that labels the photon polarization) and the last two terms
describe the $1\leftrightarrow2$ and $1\leftrightarrow3$
transitions in the rotating wave approximation, respectively.
(See Figure
\ref{fig:fig0}.) States $|2\rangle$ and $|3\rangle$ are chosen so
that no transition between them is possible (e.g., because of
selection rules). The matrix elements of the interaction
Hamiltonian read
\andy{intel}
\barr
 \phi_{\bmsub k\lambda} &=&\frac{e}{\sqrt{2\epsilon_0 V\omega}} \int
d^3x\;e^{-i\bmsub k\cdot\bmsub x}\bm\epsilon_{\bmsub
k\lambda}^*\cdot \bm j_{12}(\bm x),
\nonumber\\
\Phi_{\bmsub k\lambda}&=&\frac{e}{\sqrt{2\epsilon_0 V\omega}} \int
d^3x\;e^{-i\bmsub k\cdot\bmsub x}\bm\epsilon_{\bmsub
k\lambda}^*\cdot \bm j_{13}(\bm x),
\earr
where $-e$ is the electron charge, $\epsilon_0$ the vacuum
permittivity, $V$ the volume of the box, $\omega=|\bm k|$,
$\bm\epsilon_{\bmsub k\lambda}$ the photon polarization and $\bm
j_{\rm fi}$ the transition current of the radiating system. For
example, in the case of an electron in an external field, we have
$\bm j_{\rm fi}=\psi_{\rm f}^\dagger \bm\alpha\psi_{\rm i}$ where
$\psi_{\rm i}$ and $\psi_{\rm f}$ are the wavefunctions of the
initial and final state, respectively, and $\bm\alpha$ is the
vector of Dirac matrices. For the sake of generality we are using
relativistic matrix elements, but our analysis can also be
performed with nonrelativistic ones $\bm j_{\rm fi}=\psi_{\rm
f}^*\bm p\psi_{\rm i}/m_e$, where $\bm p/m_e$ is the electron
velocity.

We only give here the main results. The detailed calculation can
be found in \cite{PFFP}. The temporal evolution is found by
solving the time-dependent Schr\"odinger equation
\andy{Schrt}
\beq\label{eq:Schrt}
i\frac{d}{dt}|\psi(t)\rangle= H(t)|\psi(t)\rangle,
\eeq
for the states [$|i;n_{\bmsub k\lambda}\rangle$ = atom in state
$|i\rangle$ and $n_{\bmsub k \lambda}$ $(\bm k,\lambda)$-photons]
\andy{statesdefin}
\beq\label{eq:statesdefin}
|\psi(t)\rangle=\As(t)|2;0\rangle+{\sum_{\bmsub
k,\lambda}}'y_{\bmsub k\lambda}(t)|1;1_{\bmsub
k\lambda}\rangle+{\sum_{\bmsub k,\lambda}}'z_{\bmsub k\lambda}(t)
e^{-i\Omega_0 t} |3;1_{\bmsub k\lambda}\rangle
\eeq
[a prime means that the summation does not include the laser
frequency $(\bm k_0,\lambda_0)$] and initial condition
$|\psi(0)\rangle=|2;0\rangle$.

By Fourier-Laplace transforming (\ref{eq:Schrt}) and
incorporating the initial conditions, the solution reads
\andy{xs}
\beq
\wtilde \As(E)=\frac{i}{E-\omega_0-\Sigma(B,E)},
\qquad
\Sigma(B,E)=\frac{1}{2}\left[\Sigma(E+B)+\Sigma(E-B)\right],
\label{eq:xs}
\eeq
with
\andy{Q(B,s)dipdiscr}
\beq\label{eq:Q(B,s)dipdiscr}
\Sigma(E)=
\sum_{\bmsub k,\lambda}\frac{|\phi_{\bmsub
k\lambda}|^2}{E-\omega_k},
\qquad
B^2=\bar N_0\,|\Phi_{\bmsub k_0\lambda_0}|^2.
\eeq
$B^2$ is proportional to the intensity of the laser field ($\bar
N_0=$ average photon number) and can be viewed as the ``strength"
of the observation performed by the laser beam on level
$|2\rangle$ (in the sense of Section
\ref{sec-QZEcont1}).

\begin{figure}[t]
\begin{center}
\epsfig{file=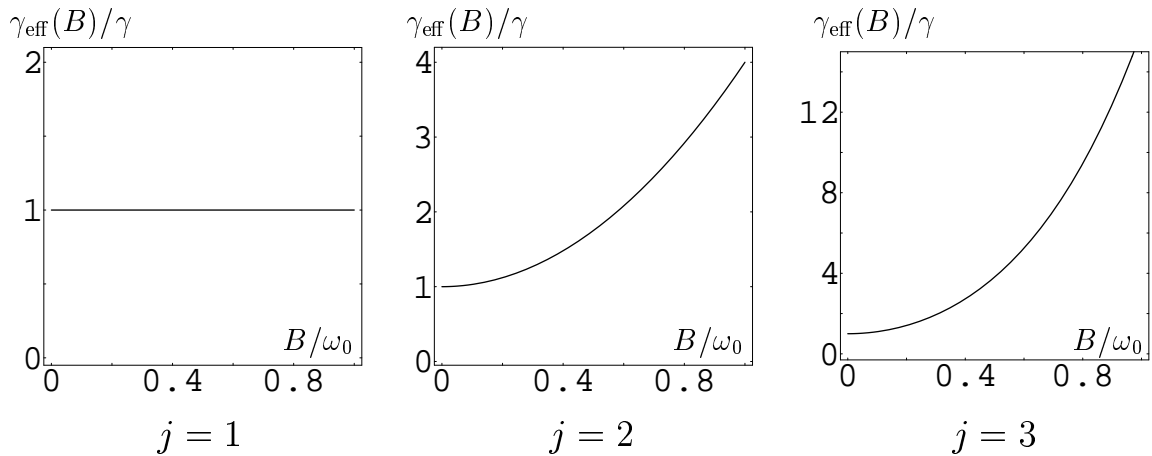,width=\textwidth}
\end{center}
\caption{The decay rate $\gamma_{\rm eff}(B)$ vs $B$, for electric
transitions with $j=1,2,3$; $\gamma_{\rm eff}(B)$ is in units
$\gamma$ and $B$ in units $\omega_0$. Notice the different scales
on the vertical axis.}
\label{fig:gamma(B)dip}
\end{figure}

The dynamics is dominated by the pole of (\ref{eq:xs}), that is
solution of the equation
\beq
E_{\rm pole} -\omega_0 -\Sigma(B, E_{\rm pole})=0,
\eeq
where  $\Sigma(B,E)$ is of order $g^2$ ($g=$ coupling constant).
The position of the pole $E_{\rm pole}$ (and as a consequence the
decay rate $\gamma=-2\,\Im\, E_{\rm pole}$) depends on the value
of $B$. One must handle with care the branch cuts arising from the
self-energy function (\ref{eq:Q(B,s)dipdiscr}): within the
convergence radius $R_c=\omega_0-B$ one obtains, for $|E_{\rm
pole}-\omega_0|<R_c$,
\andy{poloB}
\barr
E_{\rm pole} &=&\omega_0+\frac{1}{2}\left[\Sigma(\omega_0+B+i0^+)
+\Sigma(\omega_0-B+i0^+)\right]+\Ord (g^4).
\label{eq:poloB}
\earr
By writing
\andy{s(B)}
\beq\label{eq:s(B)}
E_{\rm pole}=\omega_0+\Delta(B)-i\frac{\gamma_{\rm eff}(B)}{2}
\eeq
and substituting in (\ref{eq:poloB}), after some calculations
\cite{PFFP} one obtains
\andy{MPSs}
\beq\label{eq:MPSs}
\gamma_{\rm eff}(B)= \gamma \; \frac{\Sigma(\omega_0+B)+
\Sigma(\omega_0-B)\theta(\omega_0-B)}{2\Sigma(\omega_0)} +\Ord
(g^4),
\eeq
expressing the ``new" lifetime $\gamma_{\rm eff}(B)^{-1}$, when
the system is shined by an intense laser field $B$, in terms of
the ``ordinary" lifetime $\gamma^{-1}$, when there is no laser
field. By taking into account the general behavior of the matrix
elements of the interaction \cite{PFFP,Berestetskii}, one gets to
$\Ord (g^4$)
\andy{gamma(B)dip}
\beq\label{eq:gamma(B)dip}
\gamma_{\rm eff}(B)\simeq
\frac{\gamma}{2}\left[\left(1+\frac{B}{\omega_0}\right)^{2j\mp 1}
+\left(1-\frac{B}{\omega_0}\right)^{2j\mp 1}
\theta(\omega_0-B)\right], \qquad (B \ll \Lambda)
\eeq
where $\mp$ refers to 1-2 transitions of electric and magnetic
type, respectively, $j$ is the angular momentum of the photon
emitted in the 2-1 transition and $\Lambda
\sim$ (Bohr radius)$^{-1}$ is the frequency cutoff of the
interaction, so that the case $B <
\omega_0 \ll \Lambda$ is the physically most relevant one. The
decay rate is profoundly modified by the presence of the laser
field. Its behavior is shown in Figure \ref{fig:gamma(B)dip} for
a few values of $j$. In general, for $j>1$ (1-2 transitions of
electric quadrupole, magnetic dipole or higher), the decay rate
$\gamma_{\rm eff}(B)$ increases with $B$, so that the lifetime
$\gamma_{\rm eff}(B)^{-1}$ decreases as $B$ is increased. Since
$B$ is the strength of the observation performed by the laser
beam on level $|2\rangle$, this is an IZE, for decay is {\em
enhanced} by observation.

Equation (\ref{eq:gamma(B)dip}) is valid for $B\ll\Lambda$. In
the opposite case $B\gg\Lambda$, one gets to \Ord ($g^4$)
\andy{gammaom}
\beq\label{eq:gammaom}
\gamma_{\rm eff}(B) \simeq \frac{\gamma}{2} \;
\frac{\Sigma(B)}{\Sigma(\omega_0)} \propto (B/\Lambda)^{-\beta}.
 \qquad (B \gg \Lambda)
\eeq
This result is similar to that obtained in \cite{Mihokova97}. If
such high values of $B$ were experimentally obtainable, the decay
would be hindered (QZE).

One can study several other features of this process, like the
photon spectrum, the dressed (Fano) states \cite{Fano} and some
interesting links with electromagnetically induced transparency
\cite{induced}. Additional details can be found in \cite{PFFP}.

\section{Concluding remarks}
 \label{sec-conc}
 \andy{conc}

The dynamical features of a quantum system are always modified by
the action of an external agent. In some cases, one can
conveniently regard the interaction as a sort of ``close look" at
the system. When the effect of such interaction can be accurately
described as a projection operator {\em \`a la} von Neumann, one
obtains the usual formulation of the quantum Zeno effect in the
limit of frequent measurements. Otherwise, if the description in
terms of projection operators does not apply, but one can still
properly think in terms of a ``continuous gaze" at the system, it
turns out very convenient to interpret the resulting dynamics as
a Zeno effect due to continuous measurement.

The form factors of the interaction play a primary role when the
quantum system is ``unstable." In this case, a transition from
Zeno to inverse Zeno (Heraclitus) becomes possible and a
sufficient condition can be given that characterizes the
transition between the two regimes. The inverse quantum Zeno
effect has interesting applications in other branches of physics,
and turns out to be relevant in the context of quantum chaos and
Anderson localization \cite{chaos}.

\section{Acknowledgements}
 \label{sec-ack}
 \andy{ack}
We thank H. Nakazato and L.S.\ Schulman for interesting comments.
This work is supported by the TMR-Network of the European Union
``Perfect Crystal Neutron Optics" ERB-FMRX-CT96-0057.


\end{document}